\documentclass[12pt]{article}
\usepackage{amssymb,amsmath}
\begin{document}

%%%%%%%%%%%%%%%%%%%%%%%%%%%%%%%%%%%%%%%%%%%%%%%%%%%%%%%%%%%%%%%%%%%%%
%%%%%%%%%%%%%%      TITLE PAGE     %%%%%%%%%%%%%%%%%%%%%%%%%%%%%%%%%%
%%%%%%%%%%%%%%%%%%%%%%%%%%%%%%%%%%%%%%%%%%%%%%%%%%%%%%%%%%%%%%%%%%%%%

\sloppy
\title
%{\hfill{\normalsize\sf FIAN/TD/01-15}    \\
 %           \vspace{1cm}
{\Large  On two pictures in the heuristic approach to gravity }

\author
 {
       A.I.Nikishov
          \thanks
             {E-mail: nikishov@lpi.ru}
  \\
               {\small \phantom{uuu}}
  \\
           {\it {\small} I.E.Tamm Department of Theoretical Physics,}
  \\
               {\it {\small} P.N.Lebedev Physical Institute, Moscow, Russia}
  \\
  %       {\it {\small} 117924, Leninsky Prospect 53, Moscow, Russia.}
 }
%
%--------------------------------------------------------------------
\maketitle
%--------------------------------------------------------------------
%%%%%%%%%%%%%%%%%%%%%%%%%%%%%%%%%%%%%%%%%%%%%%%%%%%%%%
\begin{abstract}
We examine the heuristic approach to constant gravitational field by Dehnen,
 H\"onl and Westpfahl, extending it everywhere  beyond linear approximation.
 Then it becomes flexible to accommodate possible modifications of General
 Relativity. We have found that two pictures introduced in the related paper by 
 Thirring are helpful in better understanding some features of gravitation.
 In particular, this approach suggest that the privileged  system for constant
 gravitational field must be the isotropic one and that the requirement of
 gauge invariance in gravitation theory may be a luxury; it is sufficient to
 take care that the nonphysical degrees of freedom do not invalidate
 calculations. It follows from this approach that gravitational constant
 should depend on gravitational field and some universality in the form of
 metric of an asymmetric body is suggested.
\end{abstract}

The concept of two pictures in considering gravitational effects was introduced
explicitly by Thirring in 1961 [1]. The essential  ingredients are contained 
already in paper [2]. We assume that the metric of the constant gravitational
 field has the form
$$
ds^2=g_{00}dt^2+g_s(dx_1^2+dx_2^2+dx_3^2),\quad 
g_s\equiv g_{11}=g_{22}=g_{33}.                           \eqno(1)
$$
Or denoting
$$
 g_{\mu\nu}=\eta_{\mu\nu}+h_{\mu\nu}, \quad \eta_{\mu\nu}={\rm diag}(-1,1,1,1,),
                                                                     \eqno(2)
 $$
 we have  
 $$
    h_{s}=h_{11}=h_{22}=h_{33},             \eqno(3)
 $$
 For a spherically symmetric body only isotropic form of the Schwarzschild 
 solution has the form (1).

 It is proposed in [2] and [1] to use as a rod and a clock, for example, the
 hydrogen atom. It is important that one and the same system do both jobs:
 if the gravitational field simply redshifted the atomic spectrum, this 
 means that the atom must retain its spherical symmetry. It is assumed that
 tidal force is negligible and gravitational field can be considered as
 constant (homogeneous) inside the atom.
  Then the metric (1) can be interpreted in the following way:
 in the presence of the gravitational field the Bohr radius $r_b$ becomes
 $r_{b\phi}$:
$$
r_b\to r_{b\phi}=\frac{r_b}{\sqrt{g_{s}}},
  \quad r_b=\frac{\hbar^2}{me^2}.                               \eqno(4) 
$$
It shrinks. Yet  we consider $r_b$ as a unit of length also in the 
gravitational field. To obtain this "true " or "observable" length from
the coordinate one $r_{b\phi}$, we multiply it by $\sqrt{g_{s}}$.
So the length $\Delta l$ in the presence of the gravitational field becomes
$$
\Delta l\to \Delta l_{\phi}=\frac{\Delta l}{\sqrt{g_{s}}}=\Delta l
(1+\phi+\cdots).                                                       \eqno(5)
$$

Similarly we treat the period of oscillation T, interval $\Delta t$
and frequency $\omega$:
$$
 T\to T_{\phi}=\frac{T}{\sqrt{|g_{00}|}},\quad 
   \Delta t\to\Delta t_{\phi}=
  \frac{\Delta t}{\sqrt{|g_{00}|}}, \quad \omega\to \omega_{\phi}=
  \omega\sqrt{|g_{00}|}=\omega(1+\phi+\cdots),
                                                                     \eqno(6)
$$
(It is said that  $\omega_{\phi}$ is measured in
the world time $t$, i.e. by the clock ticking in the flat space time from which
we start to calculate the influence of the gravitational field on the
atom.)
So the appearance of metric (1) becomes visually evident. 

It follows from (1) that the coordinate velocity of light is given by
$$
c_{\phi}^2=c^2\frac{|g_{00}|}{g_{s}}=c^2(1+4\phi+\cdots).      \eqno(7)
$$
The rest energy $mc^2$ in the gravitational field becomes
$$
mc^2\to m_{\phi}c_{\phi}^2=mc^2\sqrt{|g_{00}|}=mc^2(1+\phi+\cdots),   \eqno(8)
$$
see for example eq. (88.9) in [3].
From (7) and (8) we have
$$
m\to m_{\phi}=m\frac{g_{s}}{\sqrt{|g_{00}}|}=m(1-3\phi+\cdots),       \eqno(9)
$$
 and from (6) and (8) $mc^2T=m_{\phi}c_{\phi}^2T_{\phi}$, i.e. the quantity
 of dimension $erg\cdot sec$ is independent of $g_{\mu\nu}$. It follows
 from here  (and from $\omega T=\omega_{\phi}T_{\phi}, E=\hbar \omega$) that
 Planck constant $\hbar$ must also be independent 
of gravitational field [2] and we can put $\hbar=1$. The same should be
 true for dimensionless fine
 structure $\alpha=\frac{e^2}{\hbar c}=\frac{e^2_{\phi}}{\hbar c_{\phi}}$.
From here
$$
e^2\to e^2_{\phi}=e^2\frac{c_{\phi}}{c}=e^2\sqrt{\frac{|g_{00}|}{g_{s}}}=
e^2(1+2\phi+\cdots).
                                                                       \eqno(10)
 $$
 The dimensionless gravitational potential $\phi/c^2$ is independent of
 gravitational field ($\phi m =\frac{\phi}{c^2}mc^2$ has the dimension of
 energy; so $\frac{\phi}{c^2}$ is dimensionless.) The gravitational constant
 $G$ has the dimension $cm^3g^{-1}sec^{-2}$. Hence
 $$
  G\to G_{\phi}=G\frac{|g_{00}|^{\frac32}}{g_s^{\frac52}}=G(1+8\phi+\cdots).
                                                                      \eqno(11)
 $$
 The Einstein gravitational constant $\varkappa=\frac{8\pi G}{c^4}$ is
 independent of gravitational field only in linear approximation:
 $$
 \varkappa \to \varkappa_{\phi}=\frac{\varkappa}{\sqrt{g_s|g_{00}|}}=
 \varkappa(1+O(\phi^2)+\cdots).                                    \eqno(12)
 $$
Now, the picture in which we are dealing with quantities with
 subscript $\phi$ Thirring calls the unrenormalized one. This is the
 global picture. On the other hand, the picture, in which all the quantities
 have the same value in the gravitational field as outside of it, Thirring
 calls the renormalized picture. This is the local picture. Dealing with one
 picture it is important not to mix terms of one picture with that of
  another. Otherwise you will be in error [1].

Thirring considers the hydrogen atom starting from Minkowski space-time and
using the gravitational potentials of isotropic Schwarzschild solution
 in linear approximation
$$
 h_{\mu\nu}=-2\phi\delta_{\mu\nu},\quad \phi=-\frac{GM}{R}.     \eqno(13)
$$
 He confirms the results (4) and (6)
in the considered approximation. These results should hold beyond linear
approximation as long as the tidal forces are negligible (and in general the
influence of gravitational field on atomic spectra is very small [4], [5] ).
 To see how this comes about, we consider the Dirac equation for
the electron in hydrogen atom. In the presence of gravitational field
in global picture each quantity transforms according to rules (4)-(10),
which follow from dimensional considerations [2]. The term
$mc^2$ acquire the factor $\sqrt{g_{00}}$. The other terms, having the same
dimension, acquire the same common factor. In other words, the equations,
describing the atom, are invariant under the transformations (4-10).
 So the spectrum is simply
redshifted in this picture. The frequency $\omega_{\phi}$ is conserved in
this picture. Roughly speaking it is the total (kinetic plus potential) energy,
if we ascribe to the photon a negligible but finite mass. 

On the other hand, in the local picture in terms of observable quantities
the Dirac equation remains the same. So do the observable frequency 
$\omega=\frac{\omega_{\phi}}{\sqrt{|g_{00}|}}$. To climb up from the
gravitational potential well this photon have to lose its (kinetic) energy
 as do any  (ultrarelativistic) particle, see eq. (89.9) in [3].

 It follows from these considerations that a photon moving in gravitational
 field interacts with this field, although its frequency $\omega_{\phi}$
  remains constant. Virtual gravitons interact with the considered photon
   or atom in any coordinate system, including locally inertial
   system, although we cannot detect this interaction in the last system.
    So, in formulating the  equivalence principle, it is inaccurate 
  to say that the gravitational field is absent in the locally inertial system;
  it is simply undetectable in the small volume of space-time of that system.

  Now, the important point for Thirring's calculation  is the use of the
  isotropic coordinate system. In other coordinate system the atom will
  lose its spherical symmetry and this will distort its spectrum.
  This suggests two things. Thirst, the privileged system should be isotropic
one not harmonic as advocated by Fock [6] and Logunov [7]. Second,
  it may be a luxury to demand gauge invariance in gravitation. It may be
  sufficient to take care that the nonphysical degree of freedom do not 
  spoil calculations.

  In general relativity the isotropic system is distinguished by the fact
  that $g_{s}$ is finite for finite $r$. Yet $g_{00}$ goes to zero when 
  $r$ goes to the horizon. It seems more natural to expect that $g_{00}$
   should  be nonzero for finite $r$. We note also that $g_{00}$ and $g_s$
   of the isotropic Schwarzschild solution depends on space coordinates only
   through the Newtonian potential $\phi$. In the volume of the atom this
   field can be considered as homogeneous. The same is true
   for any constant field. So, it seems reasonable to assume that in any
   constant field the dependence of $g_{00}$ and $g_s$ on $\phi$ is the
   same, only the dependence of $\phi$ on space coordinates is determined
   by the form and other characteristics of the gravitating body.
    In other words, we
   expect that the coordinate velocity of light in any constant 
   gravitational field is given by the universal function of $\phi$,
   given by the right hand side of (7).

  Now, in the linear approximation (13), used by Thirring, the field
   satisfies the Hilbert condition
 $$
 {\bar h^{\mu\nu}}{}_{,\nu}\equiv
(h^{\mu\nu}-\frac12\eta^{\mu\nu}h)_{,\nu}=0,\quad h={h^{\sigma}}_{\sigma},\quad
h_{,\nu}\equiv\frac{\partial}{\partial x^{\nu}}h.  \eqno (14)
$$
There are reasons to believe that this condition exclude the nonphysical
degrees of freedom.
(Namely, there is a gravitational energy- momentum tensor,
which gives positive gravitational energy density of a Newtonian center
and this density can be obtained from its electromagnetic counterpart simply
  by substitution $e^2\to GM^2$ [8].)  According to Logunov beyond linear
   approximation  the  role of Hilbert condition
is played by the harmonic condition. But the harmonic coordinate system is not
isotropic. It seems strange that isotropic system does not satisfy some
easily explainable gauge condition.
 
 Assuming that Hilbert condition is consistent with isotropy beyond
the linear approximation and using it, we can relate $g_{00}$ with
 $g_s$. Then 
we obtain  $h_{00}=h_{11}=h_{22}=h_{33}$. For example, if it turns out
that $g_{00}=-exp(2\phi)$, as suggested by heuristic considerations [2],
then we get from Hilbert condition $g_s=2-exp(2\phi)$.

As mentioned above, the metric form (1) in the considered approach should be
true for any constant gravitational field, in particular for axially 
symmetric body when $g_{\mu\nu}$  depend only on two cylindrical
coordinates. In general relativity  the known solutions in this case have the
 form (1) only in
linear approximation, see eqs. (8.28) and (8.30) in [9] or eq. (13.36) in [10].
          
 \section {Acknowledgements}
 I am greatly indebted to V.I. Ritus for fruitful discussions and 
 criticism.

 The work was carried out with financial support  of Scientific Schools
 and  Russian Fund for Fundamental Research  (Grants 4401.2006.2 and 
 05-02-17217).

   \section*{References}
1. W.E. Thirring, Ann. Phys. (N.Y.) {\bf16}, 96 (1961).  \\
2. H. Dehnen, H. H\"onl, and K. Westpfahl, Ann.  der
          Phys. {\bf6}, 7 Folge, Band 6, Heft 7-8, S.670 (1960).\\
3. L.D.Landau and E.M.Lifshitz, {\sl The classical theory of
 fields}, Addison-Wesley, Cambridge, MA, 1971).\\
4. L.Parker, Phys. Rev. Lett., {\bf44}. p. 1559-1562, 1980;
 Phys. Rev. {\bf22D}, p.1922-1934 (1980). \\
 5. A. Gorbatsievich, {\sl Quantum mechanics in general relativity}, Moscow,
 URSS (2003) (in Russian).\\
6. V.A. Fock,{ \sl The Theory of Space-time and Gravitation},
New York (1964). \\
7. A.A. Logunov, Part. and Nucl. {\bf29} (1) Jan.-Feb 1998, p 1;
 {\sl The Theory of Gravitational Field}, Moscow, Nauka (2000), (in Russian).\\
8. A.I. Nikishov, Part. and Nucl., {\bf37}, pp 776-784 (2006). \\
9. J.L. Synge, {\sl Relativity: The General Theory}, Amsterdam, 1960.\\
10. P.G. Bergman, {\sl Introduction to the Theory of Relativity}, New York, 1942.

\end{document}